# Falling Coupled Oscillators & Trigonometric Sums

S.R. Holcombe

Email: srholcombe@gmail.com

A method for evaluating finite trigonometric summations is applied to a system of N coupled oscillators under acceleration. Initial motion of the nth particle is shown to be of the order $T^{2n+2}$ for small time T and the end particle in the continuum limit is shown to initially remain stationary for the time it takes a wavefront to reach it. The average velocities of particles at the ends of the system are shown to take discrete values in a step-like manner.

## 1 Introduction

***Springs and Shockwaves***

A falling tension spring can exhibit interesting and sometimes counterintuitive behaviour. This is notably observed in the 'falling slinky' where the bottom of the slinky remains stationary until a shockwave abruptly changes its velocity. As though momentarily levitating. This behaviour was examined in 1993 by Calkin in [1] and more recently in 2011 by Unruh [2] who determined the characteristics of the shockwave. Cross and Wheatland in 2012 [3] presented further refinements in modelling the slinky. These discussions have of course remained within the confines of a continuous model as their common objective is the examination of the physical behaviour of the continuous spring.

A paper by Sakaguchi in 2013 [4], however, exposes similar phenomena within the analogous but more general discrete system of coupled harmonic oscillators. In [4] several interesting features of the dynamics are noted; namely the last mass/particle remaining essentially stationary until a shockwave alters its velocity, and perhaps equally interestingly, the fact that velocities of particles behind the shockwave are approximately constant. It was also observed that the shockwaves within the discrete system have a finite width.

The numerical solution in [4] is used to make assumptions about the behaviour of the particles within the tail of the shockwave. For example, from numerical simulation it is noted/assumed that particle velocities behind the shockwave are approximately constant from which the corresponding discrete form of the Rankine-Hugoniot relation is derived. An asymptotic approximation, based upon an approximation to a continuous system, is then determined for the displacement of springs from their equilibrium lengths. An interesting question therefore is, can it be shown that the average velocity of the particles within the system, once altered by the shockwave, do in fact remain uniform? Secondly, what is the nature of these jumps in velocity caused by the shockwave, their frequency and their effect on the energy within the system? This paper will address, some of these questions, where it will be shown that the velocities of particles near the ends of the system achieve average velocities of

$$\langle v_n \rangle = \left(n + \tfrac{1}{2}\right) 2Ng\sqrt{\frac{m}{k}} \ , \ n = 0,1,2,\ldots . \qquad (1.1)$$

To do so will require examining two finite trigonometric sums.

*Trigonometric sums*

Recently Kowalenko et. al. in a series of papers determined representations for many finite trigonometric summations, either in exact form (over certain domains) or as a sum involving binomial coefficients. In 2013 Fonseca and Kowalenko [5], via a detailed analysis, determined a representation for the following summation

$$S_1(N,p) = \sum_{j=1}^{N-1} (-1)^{j+1} \sin^{2p}\left(\frac{j\pi}{N}\right). \tag{1.2}$$

In 2016, Fonseca, Glasser and Kowalenko [6], using a different method than [5], obtained various other representations for elementary trigonometric sums and showed that some of these contained the representations of Merca [7] who used yet another method in 2014. The summation in (1.2) will enable a leading term approximation of the position of the oscillators and hence provide an approximation to their initial motion. Another summation that doesn't appear in the aforementioned references (though may possibly be obtained from results in them) will enable an approximation for the average velocity of the oscillators. Namely

$$S_2(N,p) = \sum_{j=1}^{N-1} \sin\left(\frac{\pi j}{2N}\right)^{2p}. \tag{1.3}$$

The method outlined in Appendix B for obtaining representations for these summations may also be applied to obtain alternate representations to those presented in [6,7], similar in form to those in [7]. For example, using this method (1.2) may be shown to be

$$S_1(N,p) = -\frac{(-1)^{N+1}}{2} + \frac{N(-1)^{N+1}}{2^{2p}} \sum_{m=0}^{\left[\frac{p+N}{N}\right]} \binom{2p}{2Nm + \mathrm{mod}(p+N, 2N)},$$

which is a significantly different representation to that found in [5]. While this representation offers no advantage (the representation in [5] offers a reasonably simple and excellent approximation), the method by which it was obtained appears reasonably robust in its applicability. Whereas representations for trigonometric sums, specifically the basic sine related trigonometric sums, contained within [5,6,7] were arrived at using several different methods, ranging in complexity, they may all be obtained using the very simple method discussed in the appendix.

## 2. Coupled Oscillators Under Constant Acceleration

Consider N particles each of mass m, coupled by N-1 springs each having a spring constant k and each of equilibrium length $\ell_0$. The equilibrium length of the system (without acceleration) is therefore $h_0 = (N-1)\ell_0$. Let $x_n(t)$ describe the position of the nth particle. Let the end of the system have a positive coordinate and let a constant acceleration of magnitude 'g' act in the positive direction. The system of equations governing the position of each mass is therefore

$$x''_0 = g + \frac{k}{m}(x_1 - x_0 - \ell_0) + (\theta(t) - 1)\frac{r}{m}$$

$$x''_n = g - \frac{k}{m}(x_n - x_{n-1} - \ell_0) + \frac{k}{m}(x_{n+1} - x_n - \ell_0) \tag{2.1}$$

$$x''_{N-1} = g - \frac{k}{m}(x_{N-1} - x_{N-2} - \ell_0),$$

where 'r' is a restoring force, and $\theta(t)$ is a step function. In all that follows it will be convenient to introduce dimensionless variables

$$\omega_0 = \sqrt{\frac{k}{m}}, \; T_0 = \frac{1}{\omega_0} \; \text{so} \; t = T\sqrt{\frac{m}{k}}, \; x_n(t) = \frac{mg}{k}X_n(T), \; L_0 = \frac{mg}{k}\ell_0, \; h_0 = \frac{mg}{k}H_0, \; R = \frac{r}{mg}. \tag{2.2}$$

The system therefore becomes

$$X''_0 = 1 + (X_1 - X_0 - L_0) + (\theta(T) - 1)R$$

$$X''_n = 1 - (X_n - X_{n-1} - L_0) + (X_{n+1} - X_n - L_0) \tag{2.3}$$

$$X''_{N-1} = 1 - (X_{N-1} - X_{N-2} - L_0),$$

where dashes represent derivatives with respect to T. Since R must support the entire system at t=0, $R = N$. At t=0 the system is in equilibrium and hence a recurrence relationship for the initial positions is

$$X_n(0) = X_{n-1}(0) + (N-n) + L_0. \tag{2.4}$$

To see this let $L_n = X_n - X_{n-1}$ be the length of the n-th spring. At t=0 all the acceleration terms are zero and one is left with a simple recurrence, $L_1(0) = (N-1) + L_0$, $L_n(0) = L_{n-1}(0) - 1$ and $L_{N-1}(0) = 1 + L_0$. This is easily solved to yield $L_n(0) = (N-n) + L_0$, $n = 1, 2, ..., N-1$. Substituting back yields the recurrence (2.4), which again may be solved readily to obtain

$$X_n(0) = X_0(0) + nL_0 + \frac{1}{2}n(2N - n - 1). \tag{2.5}$$

For convenience $X_0(0) = 0$ will be taken. Let $H(t)$ represent the length of the system and therefore the initial length under gravity is

$$H(0) = \sum_{n=1}^{N-1}(N-n) + L_0 = \frac{1}{2}(N-1)(2L_0 + N). \tag{2.6}$$

Next, consider coordinates that measure the displacement of each spring from its equilibrium length, i.e. $\eta_{n+1}(T) = X_{n+1}(T) - X_n(T) - L_0$. The system may now be written in terms of a column vector holding all the displacements $\eta$, and a tridiagonal matrix $\mathbf{A}$,

$$\frac{d^2}{dT^2}\eta = \mathbf{A}\eta, \quad \mathbf{A} = \begin{bmatrix} -2 & 1 & 0 & \ldots & 0 \\ 1 & -2 & 1 & \ldots & 0 \\ 0 & 1 & -2 & \ldots & \ldots \\ \ldots & \ldots & \ldots & \ldots & 1 \\ 0 & 0 & \ldots & 1 & -2 \end{bmatrix}.$$

The system has boundary conditions $\eta_n(0) = (N-n), \eta_n'(0) = 0$. Consider now normal modes of vibration where $\eta_n(T) = v_n \cos(\omega T)$ so that $\eta_n'' = -\omega^2 \eta_n$. The system becomes $(\mathbf{A} + \omega^2 \mathbf{I})\eta = \mathbf{0}$. Let $\det(\mathbf{A} + \omega^2 \mathbf{I}) = u_{N-1}$. Then it is not difficult to establish

$$u_{n+2} - Du_{n+1} + u_n = 0, \text{ where } u_1 = D, U_0 = 1 \text{ and } D = \omega^2 - 2. \tag{2.7}$$

The recurrence relationship in (2.7) may be solved via generating functions by defining $u(z) = \sum_{n=0}^{\infty} u_n z^n$. Multiplying (2.7) by $z^n$ and simplifying yields upon application of the boundary conditions

$$u(z) = \frac{1}{z^2 - zD + 1} = \frac{1}{(z - \beta_+)(z - \beta_-)} = \frac{1}{(\beta_+ - \beta_-)}\left(\frac{1}{(x - \beta_+)} - \frac{1}{(x - \beta_-)}\right). \tag{2.8}$$

Note that this is in fact the generating function for the Chebyshev polynomials of the second kind which will play a role in approximating the solution in sections 4 and 5. We may utilise a geometric series and express (2.8) in the following manner

$$u(z) = \frac{1}{(\beta_+ - \beta_-)} \sum_{n=0}^{\infty} \left\{ \left(\frac{1}{\beta_-}\right)^{n+1} - \left(\frac{1}{\beta_+}\right)^{n+1} \right\} z^n, \tag{2.9}$$

where $\beta_\pm^2 - \beta_\pm D + 1 = 0 \Rightarrow \beta_\pm = \frac{D \pm \sqrt{D^2 - 4}}{2}$. Therefore $\beta_+ \beta_- = 1$ and hence

$$u_{N-1} = \frac{1}{(\beta_+ - \beta_-)} \{\beta_+^N - \beta_-^N\}. \tag{2.10}$$

Now $\det(\mathbf{A} - \lambda \mathbf{I}) = u_{N-1} = 0$ implies the conditions $\beta_+ \neq \beta_-$, and $\beta_+^N = \beta_-^N$. These conditions in turn imply $\beta_- = \beta_+ e^{i\frac{2\pi n}{N}}$, $n = 1, 2, \ldots, N-1$. Therefore

$$\left(D - \sqrt{D^2 - 4}\right) = \left(D + \sqrt{D^2 - 4}\right) e^{i\frac{2\pi n}{N}}. \tag{2.11}$$

Substituting $D - \sqrt{D^2 - 4}$ in terms of $\beta_-$ and solving for D one obtains

$$D^2 = 4\cos^2\left(\frac{\pi n}{N}\right) \Rightarrow D_\pm = \pm 2\left|\cos\left(\frac{\pi n}{N}\right)\right|. \tag{2.12}$$

Let there be an integer m so for $n=1,2,\ldots,m$, $\cos\left(\dfrac{\pi n}{N}\right) > 0$, and for $n=m+1,\ldots,N-1$, $\cos\left(\dfrac{\pi n}{N}\right) < 0$. Then there are two situations for each D and two situations for each $\beta$. In each case $\beta$ must satisfy (i) $\beta_+\beta_- = 1$ and (ii) $\beta_- = \beta_+ e^{i\frac{2\pi n}{N}}$. Consider the $n=1,2,\ldots,m$ case. Then $\beta_\pm = e^{\pm i\frac{\pi n}{N}}$, or $\beta_\pm = -e^{\mp i\frac{\pi n}{N}}$. All these solutions satisfy (i) though only the second set satisfies the (ii). A similar situation holds for the $n=m+1,\ldots,N-1$ case. Therefore for all n

$$\beta_+ = -e^{-i\frac{\pi n}{N}},\ \beta_- = -e^{i\frac{\pi n}{N}},\ D_n = -2\cos\left(\frac{\pi n}{N}\right),\ n=1,2,\ldots,N-1. \qquad (2.13)$$

The eigenvalues of the system are therefore $\omega_n^2 = 2 + D_n$. Let their corresponding eigenvectors be denoted $\mathbf{v}_n = \left[v_{1,n}, v_{2,n}, v_{3,n}, \ldots, v_{N-1,n}\right]^T$ satisfying $\left(\mathbf{A} + \omega_n^2 \mathbf{I}\right)\mathbf{v}_n = \mathbf{0}$. This set of simultaneous equations has similar properties to the one examined above. In fact, setting $v_{1,n} = 1$ it may be shown that

$$v_{j,n} = (-1)^{j-1} u_{j-1,n},\ u_{0,n} = 1, \qquad (2.14)$$

which may be written as

$$v_{j,n} = \frac{(-1)^{j-1}}{\sqrt{D_n^2 - 4}}\left\{\beta_{n,+}^j - \beta_{n,-}^j\right\}. \qquad (2.15)$$

Substituting for $\beta$ and D therefore yields

$$\omega_n = \sqrt{2\left(1-\cos\left(\frac{\pi n}{N}\right)\right)},\ v_{j,n} = \frac{\sin\left(\dfrac{\pi n}{N}j\right)}{\sin\left(\dfrac{\pi n}{N}\right)}\ n,j = 1,2,\ldots,N-1. \qquad (2.16)$$

Each displacement may be represented as a linear combination of the eigenvectors in the following manner

$$\boldsymbol{\eta} = \gamma_1 \mathbf{v_1}\cos(\omega_1 t) + \gamma_2 \mathbf{v_2}\cos(\omega_2 t) + \ldots + \gamma_{N-1}\mathbf{v_{N-1}}\cos(\omega_{N-1}t)\ \text{or}\ \eta_n(t) = \sum_{j=1}^{N-1}\gamma_j v_{j,n}\cos(\omega_j t),$$

where $\gamma_j$ are constants pertaining to superposition amplitudes. Consider therefore

$$|\mathbf{v}_n|^2 = \frac{1}{\sin\left(\dfrac{\pi n}{N}\right)^2}\sum_{j=1}^{N-1}\sin\left(\frac{\pi n}{N}j\right)^2 = \frac{N}{2\sin^2\left(\dfrac{\pi n}{N}\right)} \Rightarrow \hat{v}_{j,n} = \sqrt{\frac{2}{N}}\sin\left(\frac{\pi n}{N}j\right),$$

where $\hat{v}_{j,n}$ are the components of the unit eigenvectors. To see this, utilise a double angle formula to remove the square within the summation and then write the resulting trigonometric function as two exponentials. The resulting summations are geometric. Therefore $\boldsymbol{\eta}(0)\cdot\mathbf{v}_n = \gamma_n|\mathbf{v}_n|$ which implies $\gamma_n = \boldsymbol{\eta}(0)\cdot\hat{\mathbf{v}}_n$. From $\eta_n(0) = N-n$,

$$\gamma_j = \sqrt{\frac{2}{N}} \sum_{n=1}^{N-1} (N-n) \sin\left(\frac{\pi n}{N} j\right) = \frac{\sqrt{2N}}{4} \frac{\sin\left(\frac{\pi j}{N}\right)}{\sin^2\left(\frac{\pi j}{2N}\right)}. \tag{2.17}$$

The displacement for the nth spring from its equilibrium length is therefore given by

$$\eta_n(T) = \sum_{j=1}^{N-1} (1+s_j) U_{n-1}(s_j) \cos(\omega_j T), \quad s_j = \cos\left(\frac{\pi j}{N}\right). \tag{2.18}$$

Here the capital U defines the functions

$$U_{n-1}(s_j) = \frac{\sin\left(\frac{n\pi j}{N}\right)}{\sin\left(\frac{\pi j}{N}\right)}, \quad U_n(s_j) = \sum_{k=0}^{n} (-2)^k \frac{(n+k+1)!}{(n-k)!(2k+1)!} (1-s_j)^k = \sum_{k=0}^{n} \mu_{n,k} (1-s_j)^k \tag{2.19}$$

which are Chebyshev's polynomials of the second kind [8]. Note while the use of k is reserved for the spring constant, its occasional use in summations should provide no confusion.

Observing $\frac{d^2}{dT^2} X_n = 1 + (\eta_{n+1}(T) - \eta_n(T))$, then upon integration the displacement of the n-th particle is found to be

$$\Delta X_n(T) = \frac{1}{2} T^2 + \frac{1}{2} \sum_{j=1}^{N-1} \left(\frac{1+s_j}{1-s_j}\right) \left(1 - \cos\left(\sqrt{2(1-s_j)}\, T\right)\right) \left(U_n(s_j) - U_{n-1}(s_j)\right). \tag{2.20}$$

# 3. The Continuum Limit

As noted in the introduction one of the interesting aspects of the falling slinky is the bottom of the slinky remaining motionless during the period of time it takes for a shockwave to reach it. In [2] it was determined that for the system where collisions do not occur, the bottom of the slinky remained stationary over a time frame of $0 < t < h_0/v$ where v is the velocity of the medium. In this section this result will be re-derived in the continuum limit of the discrete system.

Switching momentarily to normal coordinates to better compare results of those found in [2], the equilibrium length of the spring is given by $h_0 = (N-1)\ell_0$. We now fix $h_0$ and let N become very large defining $h_0/\Delta y = N$, where 'y' is a variable of length measured intrinsically to the system. Taking the number of particles without limit while holding m the same, increases the mass of the system without bound. Therefore fix M, the total mass of the system, and assume a constant mass density $\rho$. This implies $\rho = M/h_0 \Rightarrow m = \rho \Delta y$. Similarly, increasing the number of springs holding the spring constant k the same, yields an infinite force applied by the springs. Therefore construct a constant $E/\Delta y = k$, expressing elasticity independent upon length (Young's modulus). Therefore

$$\frac{k}{m} = \frac{E}{\rho} \frac{1}{\Delta y^2} . \tag{3.1}$$

The initial length of the spring, for example, is given by (2.6) which upon substitution of (3.1), in the limit $\Delta y \to 0$, produces

$$h(0) = h_0 + \frac{\rho g}{2E} \Delta y^2 N(N-1) = h_0 + \frac{\rho g}{2E} \Delta y^2 \frac{\lambda_0^2}{\Delta y^2} = h_0 + \frac{\rho g}{2E} h_0^2 .$$

This result is the same as that within [2] apart from the leading $h_0$ which is due to a different choice of system orientation. From (2.20) the displacement of the last particle n=N-1 is

$$\Delta x_{N-1}(t) = \frac{1}{2} g t^2 + \frac{mg}{2k} \sum_{j=1}^{N-1} (-1)^{j+1} \frac{1+s_j}{1-s_j} \left\{ \cos\left( \sqrt{\frac{2k}{m}(1-s_j)} t \right) - 1 \right\} . \tag{3.2}$$

This is obtained by using the following identity $\sin\left(\frac{\pi(N-1)}{N}j\right) = -\cos(\pi j)\sin\left(\frac{\pi j}{N}\right)$. The series expansion $s_j = \cos\left(\pi j \frac{\Delta y}{h_0}\right) = 1 - \frac{1}{2}\left(\pi j \frac{\Delta y}{h_0}\right)^2 + O(\Delta y^4)$, and (3.1) results in

$$\lim_{N \to \infty} \Delta x_{N-1}(t) \approx \frac{1}{2} g t^2 - \frac{2g h_0^2}{v^2 \pi^2} \sum_{j=1}^{\infty} \frac{(-1)^j}{j^2} \cos\left( v \left(\frac{\pi j}{h_0}\right) t \right) + \frac{2g h_0^2}{v^2 \pi^2} \sum_{j=1}^{\infty} \frac{(-1)^j}{j^2} , \tag{3.3}$$

where $v = \sqrt{E/\rho}$, and where higher order terms have been neglected. It may be shown (see Appendix A)

$$2 \sum_{j=1}^{\infty} \frac{(-1)^j \cos\left(\frac{j\pi x}{a}\right)}{\left(\frac{j\pi}{a}\right)^2} = -\frac{1}{6} a^2 + \frac{1}{2} x^2 \qquad 0 \leq x < a , \tag{3.4}$$

the use of which provides

$$\Delta x_{N-1}(t) = \frac{1}{2}gt^2 - g\left(-\frac{h_0^2}{6v^2} + \frac{1}{2}t^2\right) - \frac{2gh_0^2}{\pi^2 v^2}\frac{\pi^2}{12}, \qquad 0 \leq t < h_0/v . \qquad (3.5)$$

The last summation in (3.3) is well known and related to the Riemann zeta function, $\sum_{j=1}^{\infty}(-1)^{j+1} j^{-2} = \pi^2/12$. Cancelling terms it is therefore found

$$\Delta x_{N-1}(t) = 0 \qquad 0 \leq t < h_0/v . \qquad (3.6)$$

That is, the last particle in the system remains stationary at least for the time it takes a wave front to reach it. Note that from (3.2) this is not the case for the discrete system. In the discrete system the final particle moves instantaneously, albeit slowly (how slowly will be discussed in the next section). This is expected as each spring connecting each particle is massless. The continuum limit implies the more particles (inertia) added to the system, the longer it will take for the initial motion of the last particle to become significant. This is especially evident given the system in the continuum limit is describing a spring with mass.

## 4. Initial Motion

In the previous section it was noted that while the end particle moves instantaneously, in the continuum limit initially it remains stationary. Adding more mass/particles to the system should therefore slow the initial motion of the end particle. In this section it will be shown that the initial motion of the nth particle is approximately $NT^{2n+2}/(2n+2)!$. To do so will require the evaluation of several finite trigonometric summations.

Consider writing the cosine term in (2.18) as a series, i.e.

$$\Delta X_n(T) = \frac{1}{2}T^2 - \frac{1}{2}\sum_{p=1}^{\infty}\frac{(-1)^p 2^p T^{2p}}{(2p)!}\sum_{j=1}^{N-1}(1+s_j)(1-s_j)^{p-1}\left(U_n(s_j) - U_{n-1}(s_j)\right). \tag{4.1}$$

Reindex the summation by letting $n = N-1-m$ and note after some simplification

$$U_{N-1-m}(s_j) - U_{N-1-m-1}(s_j) = (-1)^j \left(U_m(s_j) - U_{m-1}(s_j)\right). \tag{4.2}$$

Then from (2.19) the displacements may be written as

$$\Delta X_{N-m-1}(T) = \frac{1}{2}T^2 + \frac{1}{2}\sum_{p=1}^{\infty}\frac{(-1)^p 2^p T^{2p}}{(2p)!}\sum_{j=1}^{N-1}(-1)^{j+1}(1+s_j)(1-s_j)^{p-1}\left(\sum_{q=0}^{m}\mu_{L,q}(1-s_j)^k - \sum_{q=0}^{m-1}\mu_{L-1,q}(1-s_j)^q\right).$$

(4.3)

Swapping the order of summation and using the definitions of (9.3) within 'summation I' from Appendix B the displacements become,

$$\Delta X_{N-m-1}(T) = \frac{1}{2}T^2 + \frac{1}{2}\sum_{p=1}^{\infty}\frac{(-1)^p 2^p T^{2p}}{(2p)!}\left(\sum_{q=0}^{m}\mu_{m,q}C_1(N,p+q) - \sum_{q=0}^{m-1}\mu_{m-1,q}C_1(N,p+q)\right).$$

From (9.3) $C_1(N,1) = 1$ for which the only combination of p and q occurs when p=1,q=0. Extracting this term and evaluating the corresponding $\mu$ coefficient yields a $-T^2/2$ term, hence,

$$\Delta X_{N-m-1}(T) = \frac{1}{2}\sum_{p'=1}^{\infty}\frac{(-1)^p 2^p T^{2p}}{(2p)!}\left(\sum_{q=0}^{m}\mu_{m,q}C_1(N,p+q) - \sum_{q=0}^{m-1}\mu_{m-1,q}C_1(N,p+q)\right).$$

Dashes on the summations imply the $p+q=1$ combination is omitted. Now split the summation into two parts

$$\Delta x_{N-m-1}(T) = \frac{1}{2}\sum_{p'=1}^{N-m-1}\frac{(-1)^p 2^p T^{2p}}{(2p)!}\left(\sum_{q=0}^{m}\mu_{m,q}C_1(N,p+q) - \sum_{q=0}^{m-1}\mu_{m-1,q}C_1(N,p+q)\right)$$
$$+ \frac{1}{2}\sum_{p'=N-m}^{\infty}\frac{(-1)^p 2^p T^{2p}}{(2p)!}\left(\sum_{q=0}^{m}\mu_{m,q}C_1(N,p+q) - \sum_{q=0}^{m-1}\mu_{m-1,q}C_1(N,p+q)\right).$$

Note $C_1(N,p+q) = 0$ for $1 < p+q < N$ and hence the first summation is zero. Reindexing yields therefore

$$\Delta X_n(T) = \frac{1}{2}\sum_{p'=n+1}^{\infty}\frac{(-1)^p 2^p T^{2p}}{(2p)!}\left(\sum_{q'=0}^{N-n-1}\mu_{N-n-1,q}C_1(N,p+q) - \sum_{q'=0}^{N-n-2}\mu_{N-n-2,q}C_1(N,p+q)\right).$$

The initial motion of the nth particle may be given approximately by the leading term. Consider the case for $n \neq 0$. Note the first term of the $C_1$ function occurs when $q = N-n-1$, (the last term of the first series) for which it equals $C_1(N,N) = \dfrac{N(-1)^N}{2^{N-1}}$. Evaluating the $\mu$ coefficient then determines the leading order approximation

$$\Delta X_n(T) = \frac{NT^{2n+2}}{(2n+2)!} + O(T^{2n+4}). \tag{4.4}$$

This approximation is true for all n. For while the n=0 case needs to be treated separately due to the restriction $p+q \neq 1$, the only surviving term to leading order is again the last term on the first sum (as was the case above). A similar formula then follows which is the n=0 case within (4.4).

Figure 4.1 shows the exact solution (solid line) against the corresponding leading order approximation (dashed line) for a system $N=10$.

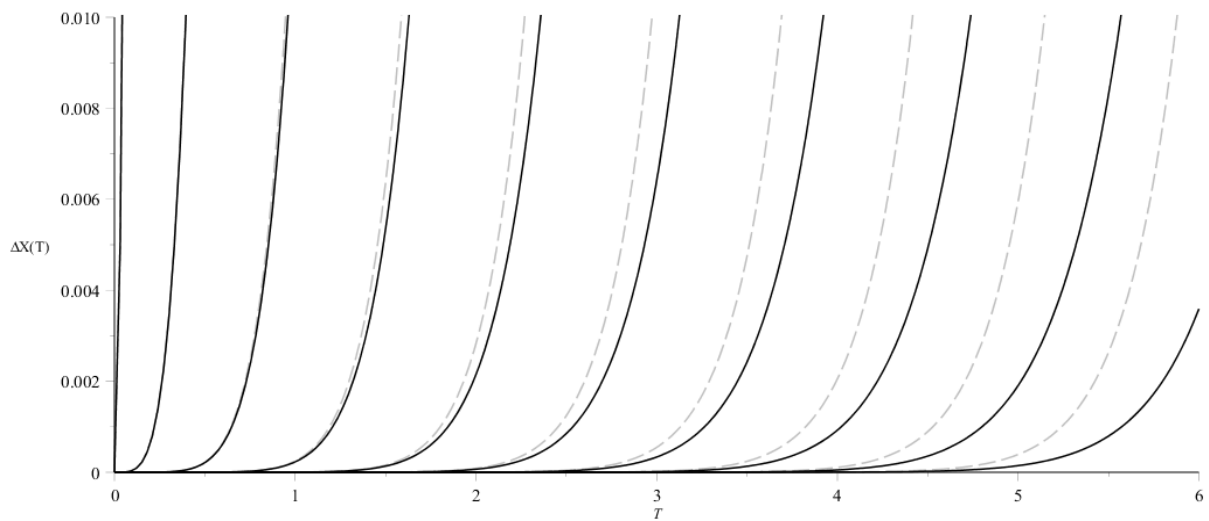

**Figure 4.1.** Initial motion of particles for a N=10 system comparing leading order estimates (dashed) against the exact solution (solid).

One might define 'significant' motion of the n-th particle to occur when the first term is O(1). That is at approximately

$$t_n \simeq \left( \frac{(2n+2)!}{Ng} \left( \frac{m}{k} \right)^n \right)^{\frac{1}{2n+2}}. \tag{4.5}$$

In any case, for small T, displacements to leading order are in a hypergeometric sequence, namely $\Delta X_{n+1}(T) = T^2 \Delta X_n(T)/(2n+4)(2n+3)$. Consider as in [4] a system of $N=10, m=1, k=1, g=.1, \ell_0=1$. Equation (4.5) estimates the motion of the end particle 'beginning' at approximately $t_9 \simeq 8.30436$ which is in accordance to the estimate of $t \sim 8$ in [4]. Figure 4.2 displays the profiles of the positions of the ten particles in this system over time.

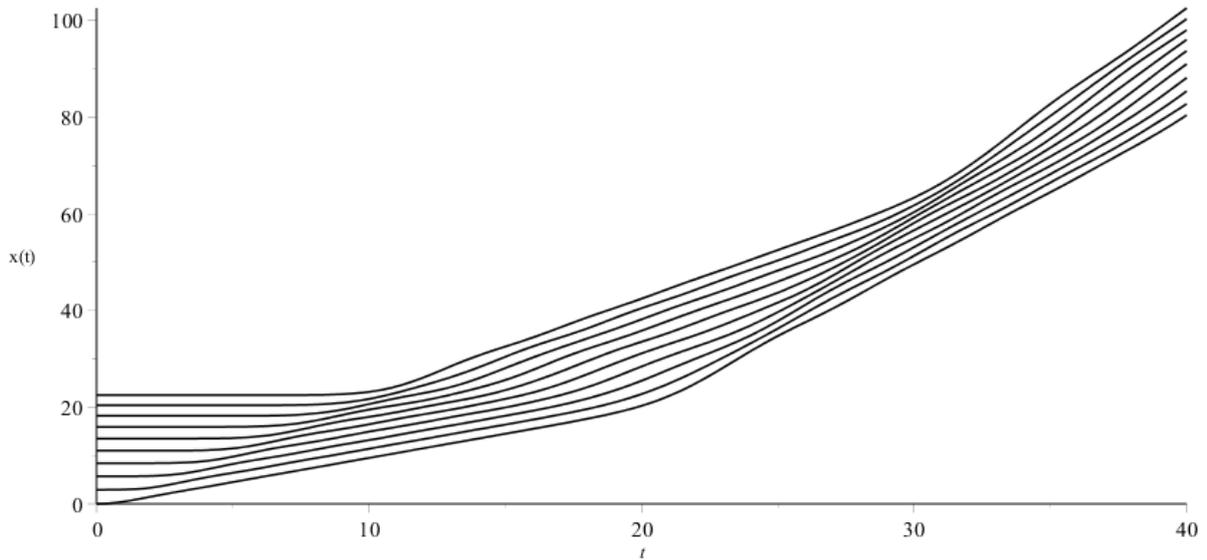

**Figure 4.2.** Position over time for a $N=10, m=1, k=1, g=.1, \ell_0=1$ system, displaying the trajectories of each particle simultaneously. The n=0, particle is the lowest most profile, the n=9 particle the upper most profile.

While each particle moves instantly at t=0, it seems reasonable to use terms such as 'delay', 'before', 'begins' when describing their motion, all the while meaning 'significant' motion (however one may want to quantify it). In figure 4.2 the delay before each particle begins moving is clearly seen, after which particles, as discussed in [4], appear to move with (approximately) uniform velocity until a shockwave once more delivers a jump in their velocities setting them again into motion with uniform velocity. This is perhaps more clearly seen in Figure 4.3 which displays the initial motion of a system consisting of $N=20, m=1, k=1, g=.1, L_0=4$.

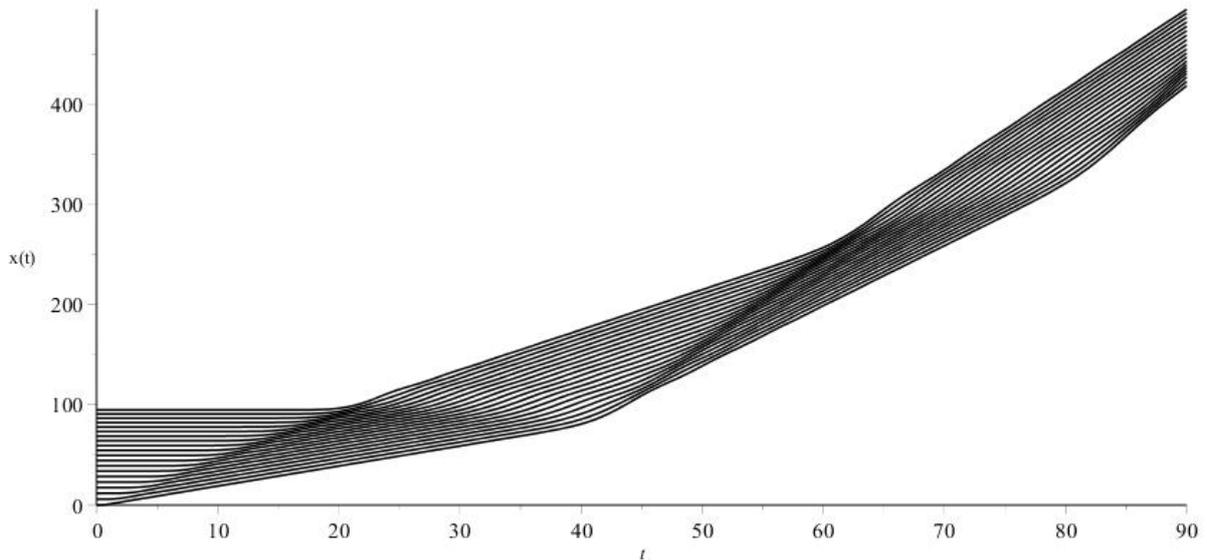

**Figure 4.3.** Position over time for a $N=20, m=1, k=1, g=.1, L_0=4$ system, displaying the trajectories of each particle simultaneously.

Note that the trajectory of the centre of mass of these systems follows that of a parabola $\frac{1}{2}T^2$, and hence the profile of all the particles will on average also adhere to that of a parabola as is illustrated in figure 4.4. While not proved here, this will be used in the subsequent section.

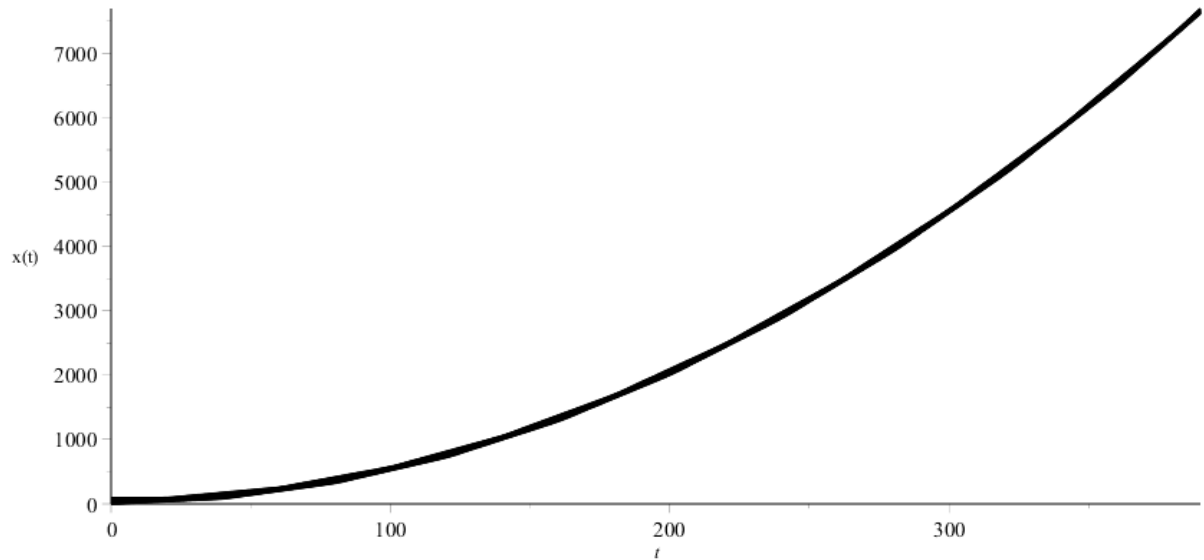

**Figure 4.4.** The $N=20, m=1, k=1, g=.1, L_0 = 4$ system over a longer time period displaying, on average, parabolic motion.

# 5 Velocities of End Particles

The step-like behaviour in velocity observed in say figure 4.3, while possibly counterintuitive, is perhaps no more astonishing than the delay occurring in the initial motion of the particles. The system is oscillatory and under constant acceleration and if there exist states of initial constant velocity then it is not unreasonable to expect to see similar states reoccurring. This section examines the step-like behaviour of the velocities of the particles where an approximation is obtained determining the average velocity for particles near the ends of the system.

In [4] it was first assumed that particles in the wake of the shockwave moved with approximately constant velocity. From this assumption the analogous Rankine-Hugoniot relations were then derived. Can it be shown, however, that the particles indeed move with near constant velocity behind the shockwave? In [4] this is somewhat achieved indirectly via a continuum limit resulting in the linearised KdV equation and the use of Airy functions approximating spring displacements from equilibrium. Here the analysis will remain within the discrete system examining particle velocities directly.

Observe for large N and particles near n=0, $U_n(s_j) - U_{n-1}(s_j) \simeq 1$. Using the series expansion for cosine, as before, the displacement of each particle is approximately (n<<N)

$$\Delta X_n(T) = \frac{1}{2}T^2 - \frac{1}{2}\sum_{p=1}^{\infty} \frac{(-1)^p 2^p T^{2p}}{(2p)!} C_2(N,p),$$

where $C_2(N,p)$ is the trigonometric summation (9.5) in Appendix B. Note that $C_2(N,1) = N-1$, and so isolating the first term and splitting the summation yields

$$\Delta X_n(T) \simeq \frac{N}{2}T^2 - \frac{N}{2}\sum_{p=2}^{2N-1} \frac{(-1)^p T^{2p}}{(2p-1)(p)!^2} - \frac{1}{2}\sum_{p=2N}^{\infty} \frac{(-1)^p 2^p T^{2p}}{(2p)!} C_2(N,p). \tag{5.1}$$

Here the value of $C_2(N,p)$ for 1<p<2N has been used. Note that the first term in (5.1) can be brought into the first summation as the p=1 term. Differentiating, the velocity for each particle for large N is therefore

$$V_n(T) \simeq -\frac{N}{2}\sum_{p=1}^{2N-1} \frac{(-1)^p 2pT^{2p-1}}{(2p-1)(p)!^2} - \Delta(T,N) . \tag{5.2}$$

Here $\Delta(T,N) = \frac{1}{2}\sum_{p=2N}^{\infty} \frac{(-1)^p 2^p T^{2p-1}}{(2p-1)!} C_2(N,p)$ and may be referred to as the tail end of the series.

Observe the following crude estimate $C_2(N,p) \leq 2^{p-1}(N-1)$, from which it may be determined

$$\Delta(T,N) \leq (N-1)\sum_{p=0}^{\infty} \frac{(2T)^{2p+4N-1}}{(2p+4N-1)!}.$$

Utilising Stirling's approximation for the factorial allows for the following bound on the tail of the series in (5.1)

$$\Delta(T,N) \leq \frac{\sqrt{N}(N-1)}{T\sqrt{2\pi}}\left(\frac{eT}{2N}\right)^{4N} \sum_{p=0}^{\infty} \left(\frac{T}{2N}\right)^{2p} . \tag{5.3}$$

The terms of this series approach zero as N increases but only for $|T|<2N$. Therefore the velocities for particles near n=0 follow approximately, for large N,

$$V_n(T) \simeq -\frac{N}{2}\sum_{p=1}^{\infty}\frac{(-1)^p 2pT^{2p-1}}{(2p-1)(p)!^2}, \qquad T<2N. \tag{5.4}$$

The series may be written in the following manner

$$V_n(T) \simeq \frac{NT}{2}\frac{\Gamma(\tfrac{1}{2})}{\Gamma(\tfrac{3}{2})\Gamma(2)}\sum_{p=0}^{\infty}\frac{(\tfrac{1}{2})_p}{(\tfrac{3}{2})_p(2)_p}\frac{(-T^2)^p}{p!} = NT\,_1F_2\left(\tfrac{1}{2};\tfrac{3}{2},2;-T^2\right), \tag{5.5}$$

where $(a)_p$ is the Pochahmmer symbol and $_pF_q$ is the generalised hypergeometric function. Observe the behaviour of $V_n(T)/N$ in figure 5.1 for the exact solution at n=0 (dashed) and the approximation (5.5).

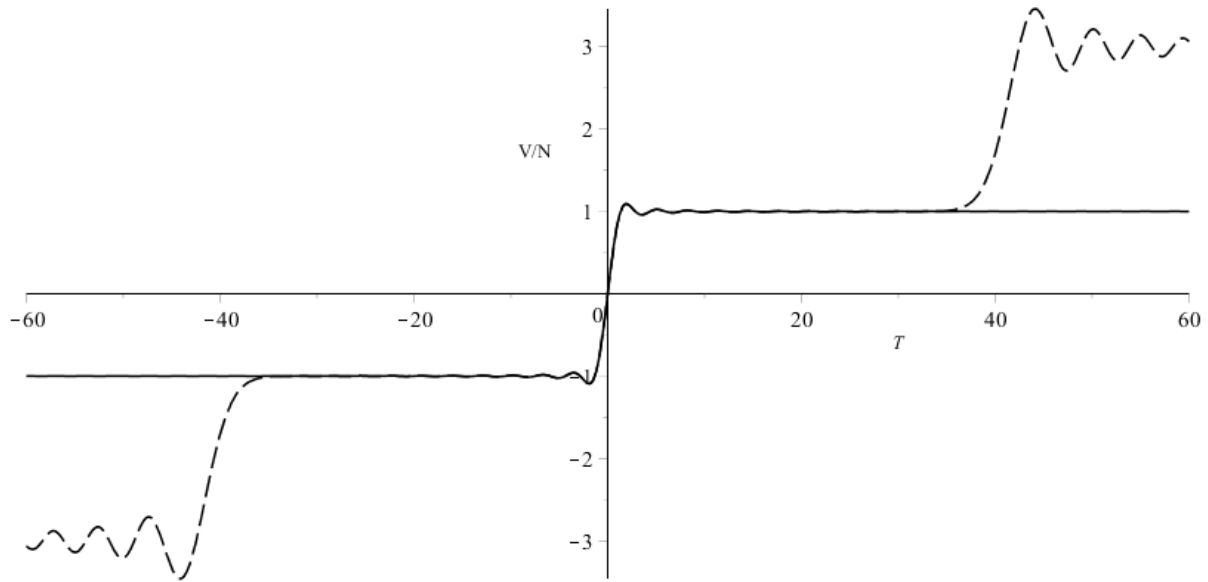

**Figure 5.1.** Scaled velocity over time for an arbitrary N system of oscillators comparing the exact solution (dashed) and approximation (solid) for the n=0 particle.

From [8, Eq. 16.11.8],

$$_1F_2\left(\tfrac{1}{2};\tfrac{3}{2},2;-x^2\right) = \frac{\Gamma(\tfrac{3}{2})\Gamma(2)}{\Gamma(\tfrac{3}{2}-\tfrac{1}{2})\Gamma(2-\tfrac{1}{2})}\frac{1}{x}+O(x^{-2}),\ x\to\infty\ \text{ hence}$$

$$\lim_{T\to\infty} V_n(T)/N = 1, \quad T<2N. \tag{5.6}$$

This result is easily determined by considering a Barnes contour representation for the hypergeometric function and then closing the contour to obtain a series in 1/x rather than x. Doing so results in the leading order expansion of (5.6). This result is independent of n. This limit is only valid, however, for particles near n=0 and for T<2N when N is large. Figure 5.2 shows the approximation (5.5) (dashed) for the scaled velocities against the exact solution (solid) for the n=0 particle and for N=5,10,15, (thin, medium, heavy).

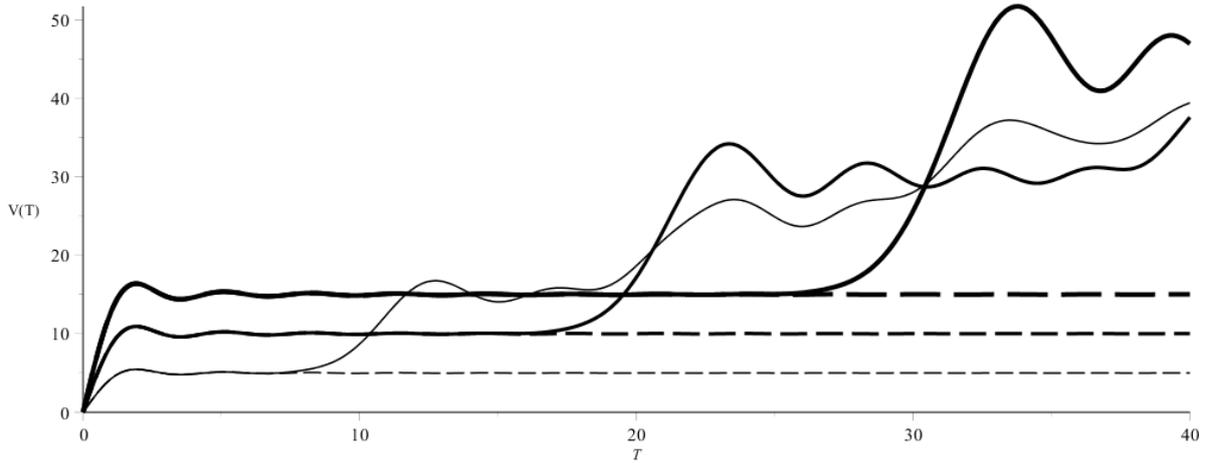

**Figure 5.2.** Velocity of the n=0 particle over time for N=5,10,15, (thin, medium, heavy) systems, comparing the approximation (dashed) to the exact solution (solid).

Note also the amplitude of the velocities on average equal N. Compare this with figure 5.3 for fixed N=100, and for n=4,19,24. Here the approximation is dashed and all exact solutions solid, left-to-right in increasing n.

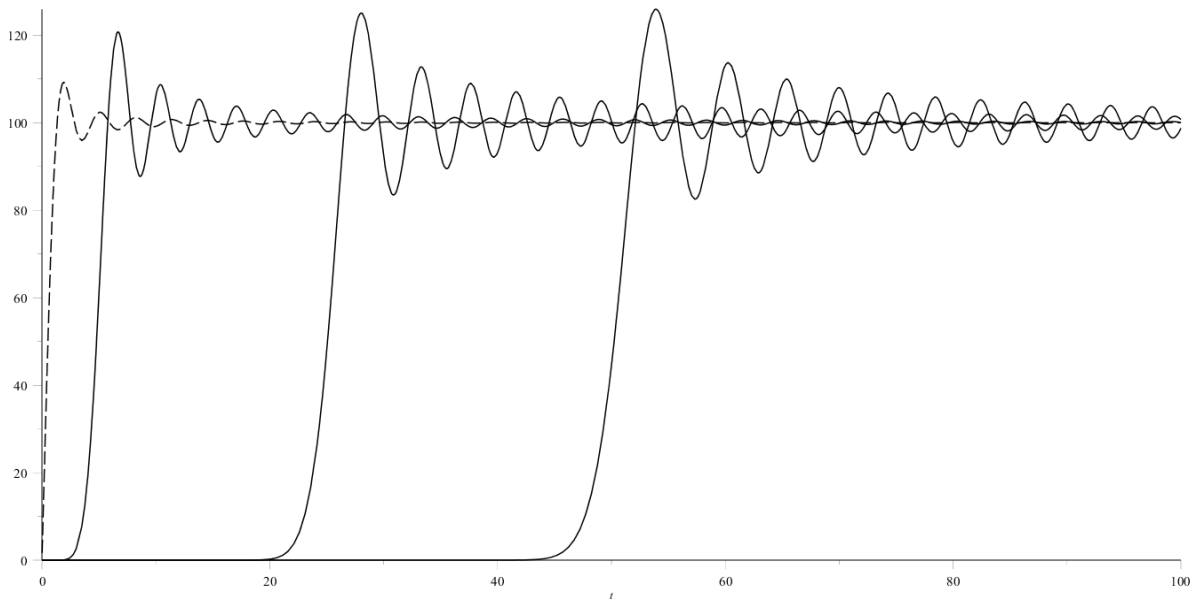

**Figure 5.3.** Velocity over time for n=4,19,24 particles within a N=100 system, comparing the n=0 approximation (dashed) to the exact solution (solid).

While the approximation fails to account for the delay in initial motion for any particle except n=0 It does, however, account for the average velocities of those particles near n=0 due to the independence of (5.6) on n.

For times greater than $2N$, the bounds on the tail of the series in (5.1) become invalid. Hence for at least a period of roughly $2N$, particles near the n=0 end of the system move at approximately the same average velocity. Reverting coordinates, where $v = x'(t)$, the limit (5.6) becomes

$$\lim_{t \to \infty} v_n(t) = gN\sqrt{\frac{m}{k}} .\qquad (5.7)$$

The result in (5.7) was derived in [4] under the assumption of constant velocity in the wake of the shockwave. Here that assumption has been shown to be true at least for particles near one end of the system over a certain time only. This result may be extended, however, to include both ends of the system and also for times beyond T=2N.

To do so, extend the definition of T to the entire real line (for diagrammatic purposes) and consider a reference frame accelerating with the system. Displacements of the nth particle within this frame are given by $\Delta \bar{X}_n(T) = \Delta X_n(T) - T^2/2$. Note that for large N, $\cos\left(\sqrt{2(1-s_j)}T\right) \simeq \cos\left(\frac{\pi j}{N}T\right)$ which yields a fundamental period for $\Delta \bar{X}_n(T)$ of approximately 2N. While displacements in this reference frame over time are not strictly periodic, for large enough N they are approximately periodic as shown in figure 5.4. Figure 5.4 shows displacements within the accelerating reference frame for N=100, where the profile for every second particle is shown.

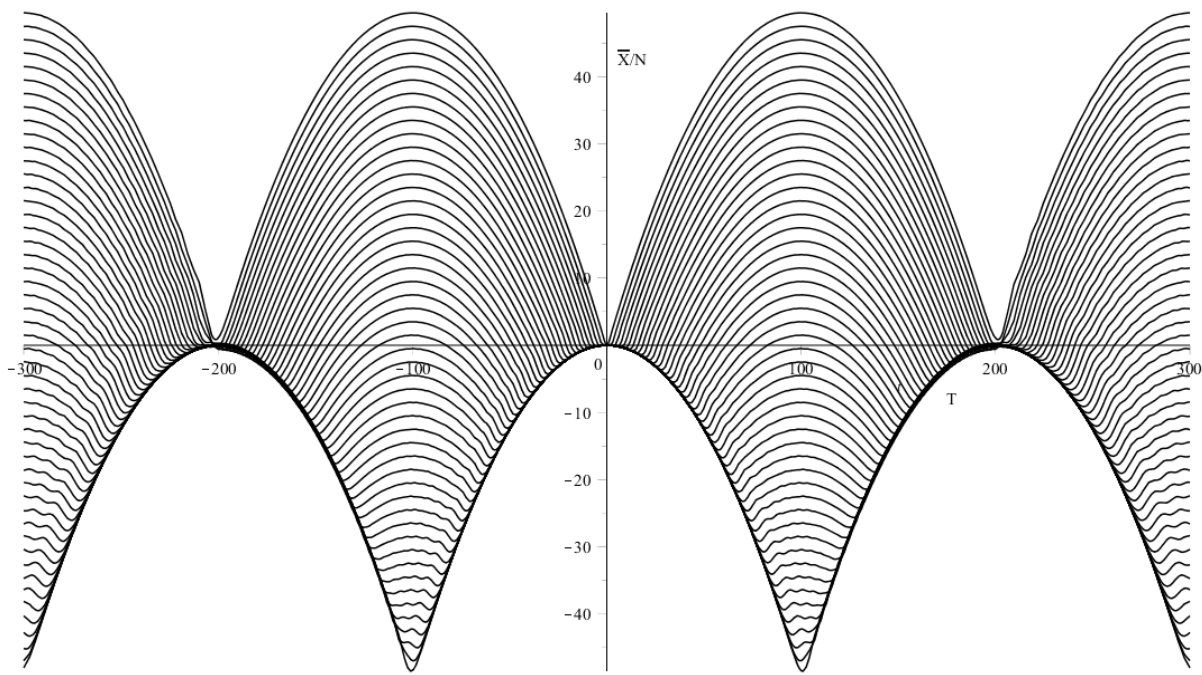

**Figure 5.4.** Scaled displacements (every second particle) over time within an accelerating reference frame for N=100. Top most and bottom most profiles are the n=0 and n=99 particles respectively.

Note the parabolic like envelope of the profiles pertaining to particles near n=0 (top most profiles) and n=N-1 particles (bottom most profiles), which appear out of phase. Observe the period of oscillation of the system as a whole is approximately 2N as previously noted.

The approximate periodicity and parabolic nature of the $\Delta \bar{X}_n(T)$ displacements carries over to step-like behaviour of the velocities in the inertial reference frame. Note in figure 5.1 the difference in velocity near T=0 is roughly $\Delta V_n(0)/N \simeq 2$. Note also the subsequent constant average velocity for a period of T=2N is given by (5.6). The aforementioned periodicity for large N implies that particles near n=0 will increase their average velocity by V=2N roughly every T=2N (via periodic extension in the accelerating reference frame). This is indeed the case as shown in figure 5.5. The dashed line indicates the velocity of an isolated particle under constant acceleration.

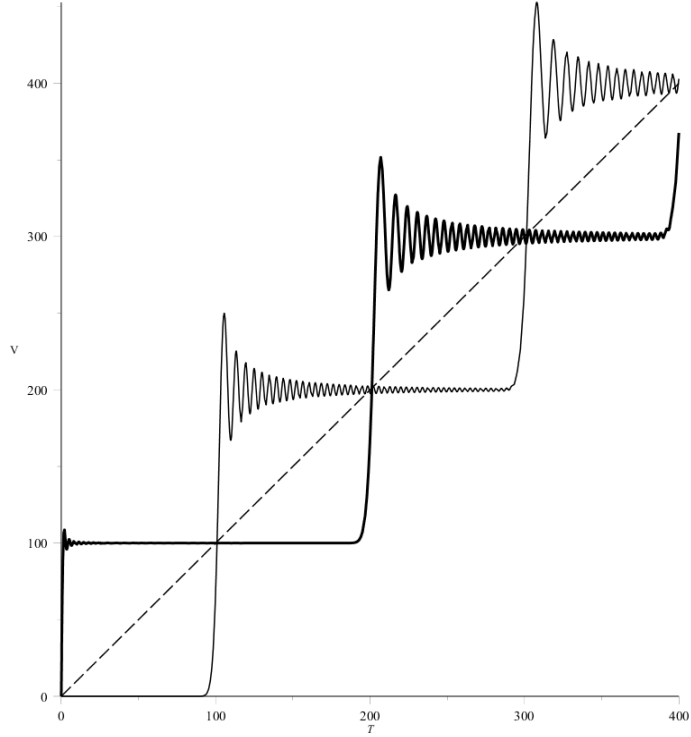

**Figure 5.5.** Velocities over time for the n=0 (heavy) and n=N-1 (fine) particles, for a N=100 system.

While, for large N, a step-like behaviour has been shown to occur for n=0 particles, this has yet to be shown for particles near n=N-1. To see that this is indeed the case, consider that for large N, particle velocities are approximately

$$V_n(T) \simeq T + \frac{1}{2}\sum_{j=1}^{N-1} \frac{\pi j}{N} \left( \frac{1+\cos\left(\frac{\pi j}{N}\right)}{1-\cos\left(\frac{\pi j}{N}\right)} \right) \sin\left(\frac{\pi j}{N}T\right) \left( \frac{\sin\left(\frac{(n+1)\pi j}{N}\right) - \sin\left(\frac{n\pi j}{N}\right)}{\sin\left(\frac{\pi j}{N}\right)} \right),$$

from which it's not difficult to establish $V_{N-1-n}(T+N) = N + V_n(T)$, for n<<N. The average velocity of the particles at both ends of the system exhibit, therefore, an out of phase step-like behaviour increasing by 2N, roughly every T=2N as shown in figure 5.5, and, for example, for $n \simeq 0$ particles given by

$$\langle V_{n\simeq 0} \rangle \simeq (k+\tfrac{1}{2})2N, \qquad k=0,1,2..., \qquad 2Nk < T < (k+1)2N.$$

A similar formula holds for $n \simeq N-1$ particles. Converting back from scaled coordinates yields the representation for the average velocities in (1.1).

## 6. Discussion

Beginning with a set of N particles, each coupled by a simple spring, the displacements of each particle was determined and shown to consist of finite trigonometric summations. The reasonably well documented phenomena of the end of the continuous system remaining stationary was observed, here via the continuum limit.

Further to this the initial motion of the particles was examined. To do so required the evaluation of the aforementioned finite trigonometric summations. Such finite trigonometric sums have gained recent attention where various different methods have been used in their evaluation. Here a single method to evaluate such summations was developed. The approach outlined in Appendix B should be noted to have significantly reduced the length/complexity of the derivation for the trigonometric sum evaluated in [5], while also providing a representation similar to that in [7]. Utilising these summations, the initial motion of the last particle, while instantaneous, was shown to be dependent upon the number of particles in the system. In fact, the initial motion of the nth particle to leading order was shown to be in a hypergeometric sequence.

Beyond the initial motion of particles, it was shown, via evaluating another finite trigonometric summation, that the average velocity of particles near the ends of the system take discrete values. Though it wasn't discussed, it should be mentioned that particles near the middle of the system also present step like behaviour in their velocities, though in a more complicated fashion where intermediate steps occur. Investigating the characteristics (as in the method of characteristics) of the continuous system would perhaps illuminate this behaviour. This, in addition to the finite width of the shockwave as observed in [4], may present interesting avenues for further investigation, as would extending (5.5) to account for a more accurate approximation dependent upon n.

## 7. Acknowledgement

I would like to express my gratitude to Victor Kowalenko for discussions regarding his recent work in the area of finite trigonometric summations, and for being very generous with his time. I also thank Edward Shin for the many useful discussions regarding the falling slinky.

# 8. Appendix A

Consider the following summation, $h(x) = \sum_{j=1}^{\infty} \frac{a}{j^2\pi^2} \cos\left(\frac{j\pi}{a}x\right)\cos\left(\frac{j\pi}{a}y\right)$, which from integration by parts twice it follows $\int_{-a}^{a} h''(x)\cos\left(\frac{j\pi}{a}x\right)dx = -\cos\left(\frac{j\pi}{a}y\right)$. This implies $h''(x) = -\delta(x-y)$. Hence $h(x) = -(x-y)H(x-y) + c_1 x + c_2$, where $c_1, c_2$ are as yet undetermined constants. Substituting this back into the integral for the coefficients yields

$$\frac{1}{a}\int_{-a}^{a}\{-(x-y)H(x-y)+c_1 x+c_2\}\cos\left(\frac{j\pi}{a}x\right)dx = c_2 + \frac{1}{2}y - \frac{a^2+y^2}{4a} + \frac{a}{j^2\pi^2}\left(\cos\left(\frac{j\pi}{a}y\right)-(-1)^j\right).$$

Letting $c_2 = -\frac{1}{2}y + \frac{a^2+y^2}{4a}$, then

$$\frac{1}{a}\int_{-a}^{a}\left\{-(x-y)H(x-y) - \frac{1}{2}y + \frac{a^2+y^2}{4a} + c_1 x\right\}\cos\left(\frac{j\pi}{a}x\right)dx = \frac{a}{j^2\pi^2}\left(\cos\left(\frac{j\pi}{a}y\right)-(-1)^j\right).$$

Consider now the sine terms for the Fourier series of $h(x)$, namely

$$\frac{1}{a}\int_{-a}^{a}\left\{-(x-y)H(x-y) - \frac{1}{2}y + \frac{a^2+y^2}{4a} + c_1 x\right\}\sin\left(\frac{j\pi}{a}x\right)dx = \frac{1}{j^2\pi^2}\left(j\pi(a-y-2ac_1)\cos(j\pi) + a\sin\left(\frac{j\pi}{a}y\right)\right)$$

Letting $c_1 = \frac{a-y}{2a}$, then

$$-(x-y)H(x-y) + \frac{1}{2}(x-y) - \frac{y}{2a}x + \frac{a^2+y^2}{4a}$$

$$= \sum_{j=1}^{\infty} \frac{a}{j^2\pi^2}\left(\cos\left(\frac{j\pi}{a}y\right)-(-1)^j\right)\cos\left(\frac{j\pi}{a}x\right) + \frac{a}{j^2\pi^2}\sin\left(\frac{j\pi}{a}y\right)\sin\left(\frac{j\pi}{a}x\right)$$

for $-a < x < a$. Using an addition formula and letting $y=x$,

$$\sum_{j=1}^{\infty}\frac{a}{j^2\pi^2}\left(1-(-1)^j\cos\left(\frac{j\pi}{\lambda_0}x\right)\right) = \frac{a^2-x^2}{4a}.$$

Noting the well-known result $\sum_{j=1}^{\infty}\frac{1}{j^2} = \frac{\pi^2}{6}$, one obtains (3.4).

# 9. Appendix B

The method for evaluating finite trigonometric sums presented here is similar to that considered in [6], though more widely applicable. Calculations in [6] hinge upon "well-known" summations in the literature. Here this dependence is avoided. This in turn allows for the determination of a summation not covered in [6,7]. It also yields representations similar in form to those in [7] for a sum analysed in [5].

*Summation I*

Before considering the first example several basic summations will be utilised and so consider,

$$\sum_{j=1}^{N-1}(-1)^{j+1}\left(1+\cos\left(\frac{\pi j}{N}\right)\right) = \sum_{j=1}^{N-1}(-1)^{j+1} + \frac{1}{2}\sum_{j=1}^{N-1}(-1)^{j+1}\left(e^{i\frac{\pi j}{N}} + e^{-i\frac{\pi j}{N}}\right)$$

$$= \frac{1}{2}\left(1+(-1)^N\right) + \frac{1}{2}\left(\frac{e^{i\frac{\pi}{N}} + (-1)^N e^{i\pi}}{1+e^{i\frac{\pi}{N}}} + \frac{e^{-i\frac{\pi}{N}} + (-1)^N e^{-i\pi}}{1+e^{-i\frac{\pi}{N}}}\right) \qquad (9.1)$$

$$= 1$$

Similarly

$$\sum_{j=1}^{N-1}\left(1+\cos\left(\frac{\pi j}{N}\right)\right) = N-1 \ . \qquad (9.2)$$

Now consider the following summation, where $p \geq 1$

$$C_1(N,p) = \sum_{j=1}^{N-1}(-1)^{j+1}\left(1+\cos\left(\frac{\pi j}{N}\right)\right)\left(1-\cos\left(\frac{\pi j}{N}\right)\right)^{p-1} = 2^p \sum_{j=1}^{N-1}(-1)^{j+1}\left(\sin\left(\frac{\pi j}{2N}\right)^{2p-2} - \sin\left(\frac{\pi j}{2N}\right)^{2p}\right).$$

The p=1 case has already been determined. Consider then

$$S_1(N,p) = \sum_{j=1}^{N-1}(-1)^{j+1}\sin\left(\frac{\pi j}{2N}\right)^{2p} = \frac{(-1)^p}{2^{2p}}\sum_{j=0}^{N-1}(-1)^{j+1}\left(e^{i\frac{j\pi}{2N}} - e^{-i\frac{j\pi}{2N}}\right)^{2p}$$

$$= \frac{(-1)^p}{2^{2p}}\sum_{j=0}^{N-1}(-1)^{j+1}\sum_{m=0}^{2p}(-1)^m\binom{2p}{m}e^{-i\frac{j\pi}{2N}m}e^{i\frac{j\pi}{2N}(2p-m)}.$$

Performing the summation over j yields

$$S_1(N,p) = \frac{(-1)^{p+1}}{2^{2p}}\sum_{m=0}^{2p}(-1)^m\binom{2p}{m}\frac{1-e^{i\pi(N+p-m)}}{1+e^{i\pi(p-m)/N}} \ .$$

Equating real components

$$S_1(N,p) = \frac{(-1)^{p+1}}{2^{2p+1}}\sum_{m=0}^{2p}(-1)^m\binom{2p}{m}\frac{\sin\left((2N-1)(N+m-p)\frac{\pi}{2N}\right)}{\cos\left((m-p)\frac{\pi}{2N}\right)} \ .$$

The summand has singularities and so care must be taken when simplifying it. Singularities occur when $p-m=(2q-1)N$, for integers q. At such points

$$\lim \frac{\sin\left((2N-1)(N+m-p)\frac{\pi}{2N}\right)}{\cos\left((m-p)\frac{\pi}{2N}\right)} = (2N-1).$$

While at all others values

$$\frac{\sin\left((2N-1)(N+m-p)\frac{\pi}{2N}\right)}{\cos\left((m-p)\frac{\pi}{2N}\right)} = (-1)^{N+m-p+1}.$$

Consider for now the case when p=N. There is a singularity at m=0, and at m=2p. Each odd multiple of N that p can take, therefore, will add two values of m where singularities occur. The total number of singularities will be even and equal to 2q. Note if $p=(2q-1)N \Rightarrow p+N=q2N$, and singularities occur at $m=2nN$ where n runs through values from 0 to $2q=\left[\frac{p+N}{N}\right]$. Suppose now, however, $(2q-1)N \leq p < (2q+1)N$, then the values of m where singularities occur become shifted by $m=2nN+L$, where $p+N-L=q2N$ that is $L=\mod(p+N,2N)=p+N-2N\left[\frac{p+N}{2N}\right]$. Therefore

$$S_1(N,p) = \frac{(-1)^{p+1}(2N-1)}{2^{2p+1}} \sum_{m=0}^{\left[\frac{p+N}{N}\right]} (-1)^{2nN+\mod(p+N,2N)} \binom{2p}{2Nm+\mod(p+N,2N)} - \frac{(-1)^{N+1}}{2^{2p+1}} \sum_{m'=0}^{2p} \binom{2p}{m}.$$

Here the summation is split over singular terms. The dashes on the last summation imply that terms where singularities occur have been omitted. We now add those terms back into the last summation the following manner,

$$S_1(N,p) = \frac{(-1)^{p+1}(2N-1)}{2^{2p+1}} \sum_{m=0}^{\left[\frac{p+N}{N}\right]} (-1)^{2nN+\mod(p+N,2N)} \binom{2p}{2Nm+\mod(p+N,2N)}$$
$$- \frac{(-1)^{N+1}}{2^{2p+1}} \sum_{m=0}^{2p} \binom{2p}{m} + \frac{(-1)^{N+1}}{2^{2p+1}} \sum_{m=0}^{\left[\frac{p+N}{N}\right]} \binom{2p}{2Nm+\mod(p+N,2N)}.$$

Note the dashes are removed. Simplifying yields

$$S_1(N,p) = -\frac{(-1)^{N+1}}{2} + \frac{N(-1)^{N+1}}{2^{2p}} \sum_{m=0}^{\left[\frac{p+N}{N}\right]} \binom{2p}{2Nm+\mod(p+N,2N)}.$$

This is a different representation than that obtained in [5].

Examining the behaviour of the falling oscillators will require use of these summations in the range $1 \leq p \leq N$. Therefore consider now the case p<N.

$$S_1(N,p) = \frac{N(-1)^{N+1}}{2^{2p}} \sum_{m=0}^{1} \binom{2p}{(2m+1)N+p} - \frac{(-1)^{N+1}}{2} = -\frac{(-1)^{N+1}}{2} \qquad p<N.$$

For p=N it is determined

$$S_1(N,N) = \frac{N(-1)^{N+1}}{2^{2N}} \sum_{m=0}^{2} \binom{2N}{2Nm} - \frac{(-1)^{N+1}}{2} = -\frac{(-1)^{N+1}}{2} + \frac{N(-1)^{N+1}}{2^{2N-1}}.$$

Hence for $1 \leq p \leq N$

$$C_1(N,p) = \sum_{j=1}^{N-1} (-1)^{j+1} (1+s_j)(1-s_j)^{p-1} = \begin{cases} 1 & p=1 \\ 0 & 1<p<N, \\ N(-1)^N/2^{N-1} & p=N \end{cases} \qquad (9.3)$$

where $s_j = \cos\left(\frac{\pi j}{N}\right)$. For p>N values of the function will simply be represented as $C_1(N,p)$.

By utilising the floor function representation of the mod function this representation can be brought into line with those of [7] where the summation runs over negative and positive terms. This effectively removes the mod/floor function from the summation. First break the summation into two parts,

$$S_1(N,p) = -\frac{(-1)^{N+1}}{2} + \frac{N(-1)^{N+1}}{2^{2p}} \sum_{m=0}^{\left[\frac{p+N}{2N}\right]} \binom{2p}{p+\left(2m+1-2\left[\frac{p+N}{2N}\right]\right)N}$$
$$+ \frac{N(-1)^{N+1}}{2^{2p}} \sum_{m=\left[\frac{p+N}{2N}\right]+1}^{\left[\frac{p+N}{N}\right]} \binom{2p}{p+\left(2m+1-2\left[\frac{p+N}{2N}\right]\right)N}.$$

In the first summation re-index $m \to \left[\frac{p+N}{2N}\right] - m$. This cancels the floor function. Then note the identity $[m/n] = 2[m/2n] + (1-(-1)^{[m/n]})/2$ [6]. Using this in the second summation and performing another re-indexing results in

$$S_1(N,p) = -\frac{(-1)^{N+1}}{2} + \frac{N(-1)^{N+1}}{2^{2p}} \sum_{m=0}^{\left[\frac{p+N}{2N}\right]} \binom{2p}{p+(-2m+1)N}$$
$$+ \frac{N(-1)^{N+1}}{2^{2p}} \sum_{m=1}^{\left[\frac{p+N}{2N}\right]+\left(1-(-1)^{\left[\frac{p+N}{N}\right]}\right)\frac{1}{2}} \binom{2p}{p+(2(m)+1)N}$$

Extracting the last term in the second summation then results in

$$S_1(N,p) = -\frac{(-1)^{N+1}}{2} + \frac{N(-1)^{N+1}}{2^{2p}} \binom{2p}{p+\left(\left[\frac{p+N}{N}\right]+\frac{1}{2}\left(1-(-1)^{\left[\frac{p+N}{N}\right]}\right)+1\right)N} + \frac{N(-1)^{N+1}}{2^{2p}} \sum_{m=-\left[\frac{p+N}{2N}\right]}^{\left[\frac{p+N}{2N}\right]} \binom{2p}{p+(2m+1)N}$$

*Summation II*

Consider the following summation

$$C_2(N,p) = \sum_{j=1}^{N-1}\left(1+\cos\left(\frac{\pi j}{N}\right)\right)\left(1-\cos\left(\frac{\pi j}{N}\right)\right)^{p-1} = 2^p \sum_{j=1}^{N-1}\left(\sin^{2p-2}\left(\frac{\pi j}{2N}\right) - \sin^{2p}\left(\frac{\pi j}{2N}\right)\right).$$

The p=1 case has been covered in (9.2). Following the same process in Summation I results in

$$S_2(N,p) = \sum_{j=1}^{N-1}\sin\left(\frac{\pi j}{2N}\right)^{2p} = \frac{(-1)^p}{2^{2p+1}}\sum_{m=0}^{2p}(-1)^m\binom{2p}{m}\frac{\sin\left((2N-1)(m-p)\frac{\pi}{2N}\right)}{\sin\left((m-p)\frac{\pi}{2N}\right)}.$$

Firstly consider the terms when $p-m=q2N$, $q=0,1,2,3,...$

$$\lim\frac{\sin\left((2N-1)(m-p)\frac{\pi}{2N}\right)}{\sin\left((m-p)\frac{\pi}{2N}\right)} = (2N-1).$$

All other terms are well defined. That is observe for $p-m \neq q2N$ we have after simplification

$$\frac{\sin\left((2N-1)(m-p)\frac{\pi}{2N}\right)}{\sin\left((m-p)\frac{\pi}{2N}\right)} = (-1)^{m-p+1}.$$

Consider the case when $p<2N$, then the only singularity occurs when m=p. When $p\geq 2N$ singularities are introduced at every integer multiple of 2N, as well as the singularity at m=p. There are therefore an odd number of singularities. In fact if $q2N \leq p < (q+1)2N$ there are 2q+1 singularities. That is the total number of singularities is $2\left[\frac{p}{2N}\right]+1$. Now consider the example when p=q2N. In this case the singularities occur at m=n2N where n runs from 0 to $2\left[\frac{p}{2N}\right]$. However if $q2N \leq p < (q+1)2N$, then the p=m singularity is shifted from an even multiple of 2N as are all the other singularities. Hence singularities occur at m=n2N+L, where $p-L=q2N$, that is $L = \mathrm{mod}(p,2N) = p - 2N\left[\frac{p}{2N}\right]$. Separating out the terms involving singularities from the summation produces

$$\sum_{j=1}^{N-1}\sin\left(\frac{\pi j}{2N}\right)^{2p} = \frac{(-1)^{p+L}}{2^{2p+1}}(2N-1)\sum_{m=0}^{2\left[\frac{p}{2N}\right]}\binom{2p}{2Nm+L} - \frac{1}{2^{2p+1}}\sum_{m'=0}^{2p}{}'\binom{2p}{m}.$$

As before, the dashes in the second summation imply omitting those terms counted in the first summation. Adding those terms back and simplifying yields

$$\sum_{j=1}^{N-1}\sin\left(\frac{\pi j}{2N}\right)^{2p} = -\frac{1}{2} + \frac{N}{2^{2p}}\sum_{m=0}^{2\left[\frac{p}{2N}\right]}\binom{2p}{2Nm+\mathrm{mod}(p,2N)}. \qquad (9.4)$$

The result (9.4) doesn't appear to be included in [6] or [7]. We may simplify this representation by utilising the floor function representation of the mod function. Substituting and splitting the summation over two ranges yields

$$\sum_{j=1}^{N-1}\sin\left(\frac{\pi j}{2N}\right)^{2p} = -\frac{1}{2} + \frac{N}{2^{2p}}\sum_{m=0}^{\left[\frac{p}{2N}\right]}\binom{2p}{2Nm+p-2N\left[\frac{p}{2N}\right]} + \frac{N}{2^{2p}}\sum_{m=\left[\frac{p}{2N}\right]+1}^{2\left[\frac{p}{2N}\right]}\binom{2p}{2Nm+p-2N\left[\frac{p}{2N}\right]}.$$

Reverse the indexing in the first sum then provides a representation in a similar form to those in [7], namely

$$\sum_{j=1}^{N-1}\sin\left(\frac{\pi j}{2N}\right)^{2p} = -\frac{1}{2} + \frac{N}{2^{2p}}\sum_{m=-\left[\frac{p}{2N}\right]}^{\left[\frac{p}{2N}\right]}\binom{2p}{p+2Nm}.$$

Again the main interest in these sums will be in the range $1 \leq p \leq N$. Here however it is perhaps more convenient to examine $1 \leq p \leq 2N$. Observe

$$2^p \sum_{j=1}^{N-1}\left(\sin^{2p-2}\left(\frac{\pi j}{2N}\right) - \sin^{2p}\left(\frac{\pi j}{2N}\right)\right)$$

$$= \frac{N}{2^p}\left(4\binom{2p-2}{p-1} - \binom{2p}{p}\right) = \frac{N}{2^p(2p-1)}\frac{(2p)!}{(p!)^2}, \qquad 1 < p < 2N$$

Hence for $1 \leq p < 2N$

$$C_2(N,p) = \sum_{j=1}^{N-1}(1+s_j)(1-s_j)^{p-1} = \begin{cases} N-1 & p=1 \\ \dfrac{N}{2^p(2p-1)}\dfrac{(2p)!}{(p!)^2} & 1 < p < 2N \end{cases} \qquad (9.5)$$

As for the previous summation, for $p \geq 2N$ values of the function will be denoted $C_2(N,p)$.

*Summation III*

Included here, as a last example, is a statement of the sine summation given in [6] and in [7], in a representation furnished via this method.

$$\sum_{j=1}^{N-1}\sin\left(\frac{\pi j}{N}\right)^{2p} = N\frac{(-1)^p}{2^{2p}}\sum_{m=0}^{2\left[\frac{p}{N}\right]}(-1)^{mN+\text{mod}(p,N)}\binom{2p}{Nm+\text{mod}(p,N)}.$$

Using the floor function representation of the mod and splitting the summation into two, produces the same result as that obtained in [7], namely

$$\sum_{j=1}^{N-1}\sin\left(\frac{\pi j}{N}\right)^{2p} = \frac{N}{2^{2p}}\sum_{m=-\left[\frac{p}{N}\right]}^{\left[\frac{p}{N}\right]}(-1)^{mN}\binom{2p}{Nm+p}.$$

# References

[1] M. G. Calkin, Motion of a falling spring, Am. J. Phys. **63**, 261 (1993)

[2] W. G. Unruh, The falling slinky, arXiv:1110.4368v1 (2011)

[3] R. C. Cross & M. S. Wheatland, Modelling a falling slinky, Am. J. Phys. **80**, 1051 (2012)

[4] H. Sakaguchi, Shockwaves in falling coupled harmonic oscillators, arXiv:1307.6686 (2013)

[5] C. M. da Fonseca & V. Kowalenko, On a finite sum with powers of cosines, Appl. Anal. Discret. Math. **7**, 354-377 (2013)

[6] C. M. da Fonseca, M. L. Glasser & V. Kowalenko, Basic trigonometric power sums with applications, Ramanujan J (2016). doi:10.1007/s11139-016-9778-0

[7] M. Merca, On some power sums of sine or cosine, Amer. Math. Monthly **121(3),** 244-248 (2014)

[8] F. W. J. Olver, A. B. Olde Daalhuis, D. W. Lozier, B. I. Schneider, R. F. Boisvert, C. W. Clark, B. R. Miller, & B. V. Saunders, eds., Digital Library of Mathematical Functions. http://dlmf.nist.gov/, Release 1.0.13 of (2016).